Title: **Can visual information encoded in cortical columns be decoded from magnetoencephalography data in humans?**

Radoslaw Martin Cichy[1], Dimitrios Pantazis[2]

[1] Computer Science and Artificial Intelligence Laboratory, MIT, Cambridge, MA, USA

[2] McGovern Institute for Brain Research, MIT, Cambridge, MA, USA

**CORRESPONDING AUTHOR**

Radoslaw Martin Cichy

Computer Science and Artificial Intelligence Laboratory

MIT

32-D430

Cambridge, MA, USA

Phone: +1 617 253 1428

Email: rmcichy@mit.edu

## ABSTRACT

It is a principal open question whether noninvasive imaging methods in humans can decode information encoded at a spatial scale as fine as the basic functional unit of cortex: cortical columns. We addressed this question in five magnetoencephalography (MEG) experiments by investigating the encoding of a columnar-level encoded visual feature: contrast edge orientation. We found that MEG signals contained orientation-specific information as early as ~50ms after stimulus onset even when controlling for confounds, such as overrepresentation of particular orientations, stimulus edge interactions, and global form-related signals. Theoretical modeling confirmed the plausibility of this empirical result. An essential consequence of our results is that information encoded in the human brain at the level of cortical columns should in general be accessible by multivariate analysis of electrophysiological signals.

# 1 INTRODUCTION

The basic format in which the primary visual cortex (V1) represents the visual world is the orientation of contrast edges (Hubel and Wiesel, 1959, 1968). Invasive research in animals and ultrahigh-resolution fMRI in humans have shown that neurons tuned to a particular edge orientation cluster in sub-millimeter size columns (Bartfeld and Grinvald, 1992; Vanduffel et al., 2002; Yacoub et al., 2008). Thus, the size of orientation columns is smaller than the sampling resolution of standard fMRI (3mm) and magnetoencephalography (MEG), seemingly raising a barrier to resolving orientations from brain data obtained with standard noninvasive techniques. However, ten years ago two studies claimed to have crossed this boundary using standard resolution fMRI (Haynes and Rees, 2005; Kamitani and Tong, 2005), showing that grating orientation can be decoded from fMRI activation patterns.

This claim has sparked a debate and has been challenged in at least two ways (Mannion et al., 2009, 2010; Kriegeskorte et al., 2010; Freeman et al., 2011, 2013; Alink et al., 2013; Carlson, 2014; Carlson et al., 2014; Pratte et al., 2014). First, interpretation of fMRI results is confounded by the complex relationship between neuronal activity, the BOLD response, and the voxel-wise sampling of BOLD activity (Logothetis and Wandell, 2004). For example, modeling voxels as compact kernels or complex spatiotemporal filters greatly influences the sampling of columnar level activity (Kriegeskorte et al., 2010), and thus complicates the interpretation of the spatial scale of the underlying signal sources.

Second, it has been argued that orientation stimuli used to probe V1 activity might also elicit orientation-specific coarse-scale activation patterns far above the size of orientation columns. Such coarse-scale activation patterns might result from a relatively stronger representation for particular orientations (Pettigrew et al., 1968; Maffei and Campbell, 1970; Mansfield, 1974; Rose and Blakemore, 1974; Kennedy and Orban, 1979; Furmanski and Engel, 2000; Li et al., 2003; Sasaki et al., 2006; Mannion et al., 2009; Freeman et al., 2011, 2013; Alink et al., 2013), boundary interaction effects between background and stimulus (Carlson, 2014), and perceptual binding processes influenced by the global form of the stimulus (Alink et al., 2013).

Here, we circumvent the first challenge by taking an alternative approach: we used MEG instead of fMRI to resolve orientation from brain signals. MEG measures direct neuronal activation without the complex convolution of the BOLD response, and its fine temporal resolution enables us to dissociate the contribution of early first-pass visual responses from late processing along the ventral visual pathway and other feedback signals. To address the second challenge, we used multiple sets of controlled stimuli to investigate alternative hypotheses of coarse-scale confounds in orientation decoding. Finally, we conducted a modeling experiment to examine whether simulated activation patterns in V1 at the spatial scale of orientation columns are discriminable from modeled MEG signals.

We found that MEG signals contained orientation-specific information even when controlling for known stimulus-induced coarse-scale activation confounds. Modeling demonstrated the theoretical feasibility of discriminating cortical activation patterns that differ at the spatial scale of cortical columns. Together, our results indicate that orientation encoding in humans at the level of cortical columns is directly accessible to experimental investigation using noninvasive electrophysiological methods. This suggests that other information encoded in the human brain in fine-grained distributed activation patterns is accessible by multivariate analysis of electrophysiological data.

## 2 MATERIALS AND METHODS

### 2.1 Participants

Experiments 1 to 5 included 12, 13, 16, 12, and 12 right-handed, healthy volunteers with normal or corrected-to-normal vision, respectively. Some subjects participated in more than one experiment, with the overall population being 20 males and 25 females, with mean age ± s.d. = 23.68 ± 4.55. The study was conducted according to the Declaration of Helsinki and approved by the local ethics committee at Massachusetts Institute of Technology.

### 2.2 Stimulus material

The stimulus set comprised diverse visual stimuli differing in local orientation and overlaid on a uniform gray background. Stimuli and background were of equal mean luminance. All stimuli were created by in-house scripts using Matlab (2014a, The MathWorks, Natwick, MA, USA).

We used two main categories of stimuli: Cartesian gratings (Figs. 1A, 2A) and logarithmic spirals (Fig. 3A). These stimuli allowed us to address the influence of different biases on orientation decoding. Cartesian gratings had frequency of 2 cycles per degree visual angle and multiple orientations (0-150° in 30° steps, and +–45°, measured with respect to the vertical line). Logarithmic spirals had 20 contrast cycles and were construed such that their edges were at an angle +45° (turning direction clockwise) or -45° (turning direction anti-clockwise) relative to a line through the center of the stimulus (the radial line) at any position in the stimulus, resulting in a local orientation disparity of 90° with each other. Both gratings and spirals were constructed in two phase exemplars with a half cycle phase shift.

For experiments 1-3, gratings and spirals were presented in an annulus with an outer radius of 10° and an inner radius of 1°.

For experiment 4, the annulus of the grating stimuli was deformed into ellipsoidal shapes. In detail, we compressed the annulus in the orientation of the grating and elongated it in the orthogonal direction. Deformations were 2, 5, 10 and 20% of the radius of the annulus. Instead of 2 phase exemplars, we completely randomized phase because we found no phase effect in experiments 1-3. A Matlab script encoding the perfect ice cube model as in (Carlson, 2014), generating the deformed stimuli, and testing for effects of cross-classification based on the ice cube model output is available upon request.

For experiment 5, we introduced patch-swapped variants for both gratings and spirals in addition to the original intact stimuli. Patch-swapped stimuli preserved the 90° disparity of the original stimuli, but were equated in global form (Fig. 5A,B) (Alink et al., 2013). To achieve this, stimuli were subdivided in patches using a polar checkerboard array defined by equi-length arcs in each of 4 concentric circles (1°, 2.5°, 5°, 10°). We created patch-swapped stimuli by swapping half (non-adjacent) patches between opposing stimuli (+45° vs. -45° gratings, or clockwise vs. counter-clockwise spirals). We repeated this procedure for each phase separately. To control for the additional edges introduced by patch-swapping, lines of background color covered the patch edges for all stimuli.

## 2.3 Experimental design

Visual stimuli were presented using Psychtoolbox (www.psychtoolbox.org) (Brainard, 1997). Stimuli appeared in random order for 0.1s, with an ISI of 0.9-1.1s. Participants were instructed to fixate on a centrally presented red fixation cross, and press a button and blink their eyes in response to a target image (displaying concentric circles) shown every 3-5 trials, to maintain attention and avoid contamination of experimental conditions with eye blink artifacts. Target image trials were not included in further analysis. The order of stimulus conditions was randomized.

Experiment 1 had 4 conditions (grating stimuli +45°/-45°, in 2 exemplar phases). Experiment 2 had 12 conditions (grating stimuli 0° to 150° in 30° steps, in 2 exemplar phases). Experiment 3 had 4 conditions (logarithmic spirals clockwise/anti-clockwise, in 2 exemplar phases). Experiment 4 had 10 conditions (grating stimuli, +45°/-45°, one original and 4 deformations to ellipsoidal shape) with each stimulus having randomized phase per trial. Experiment 5 had 8 conditions (grating stimuli +45°/-45° and spiral stimuli clockwise/anti-clockwise, original and patch-swapped versions).

In each experiment, participants completed 15 runs, each lasting 216, 255, 216, 208, and 307 s for experiments 1-5, respectively. Each experimental condition was presented 33 (Exp. 1, 2) or 13 (Exp. 3, 4, 5) times in each MEG run.

## 2.4 Human MEG recording

MEG signals were recorded from 306 channels (204 planar gradiometers, 102 magnetometers, Elekta Neuromag TRIUX, Elekta, Stockholm) at a sampling rate of 1000Hz and band-pass filtered between 0.03 and 330Hz. Raw data was pre-processed with spatiotemporal filters using Maxfilter software (Elekta, Stockholm) and then analyzed with Brainstorm (Tadel et al., 2011). We extracted peri-stimulus MEG data from -100 to +900ms with respect to each stimulus onset, and for every trial we removed the baseline (-100 to 0ms) mean of each channel, normalized with the baseline standard deviation, and temporally smoothed with a 20ms sliding window.

## 2.5 Multivariate pattern classification on MEG data

To determine the time course with which MEG signals distinguish between experimental conditions, data was subjected to multivariate pattern classification analyses using linear support vector machine (SVM) classifiers (libsvm implementation, www.csie.ntu.edu.tw/~cjlin/libsvm, Müller et al., 2001). All analyses shared a common framework: For each time point $t$ (from 100ms before to 900ms after image onset with 1ms step), single-trial MEG data were arranged in 306 dimensional pattern vectors, representing the activity in the 306 MEG sensors. To reduce computational load and improve SNR, single-trial MEG vectors were sub-averaged in groups of 40 with random assignment, yielding M averaged trials per time point and condition (M varied per experiment and type of decoding). We then measured the performance of the SVM classifier to discriminate between every pair of conditions using a leave-one-out approach: M - 1 vectors were randomly assigned to the training test, and the left-out vector to the testing set to evaluate the classifier decoding accuracy. The above procedures were repeated 100 times, each with random assignment of the raw pattern vectors to M averaged pattern vectors, and the resulting decoding accuracy was averaged over repetitions. This produced a single decoding accuracy value for each pair of conditions and each time point $t$.

In detail, we conducted 5 different classification analyses.

*Classification of orientation (same phase)*: Training and testing sets had trials with same phase. First, the SVM classifier was trained to distinguish orientation based on trials of one stimulus phase, and tested with left-out trials with the same stimulus phase. Then, the analysis was repeated for the other phase and results were averaged.

*Classification of orientation (different phase)*: Training and testing sets had trials with different phase. First, the SVM classifier was trained to distinguish orientation with trials associated with one stimulus phase, and tested for the other. Then, the analysis was repeated for the opposite arrangement of phase and results were averaged. Training was performed using the same M-1 trials as the *same phase* condition to allow a direct comparison.

*Classification of orientation (any phase)*: Training and testing sets had trials with any phase. Trials with the same orientation but 2 phase exemplars were combined, resulting in conditions with twice as many trials. The SVM classifier was trained to distinguish orientation following the previously described M-1 leave-one-out approach, but with twice greater M, and thus improved signal to noise ratio.

*Classification of phase (same orientation)*: Similar to classification of orientation (same phase), but with opposite role for the two stimuli properties.

*Classification of orientation with different stimuli shapes*: Training set had trials with grating stimuli of circular shape. Testing set had trials with the same grating stimuli of circular shape, or deformed grating stimuli of elliptical shape (with different amount of deformation from a circle). The SVM classifier was trained to distinguish orientation based on trials of the original grating stimuli, and tested for trials with the original or deformed stimuli. Phase was randomized per trial for these stimuli and was irrelevant for the classification procedure.

## 2.6 Modeling and classifying MEG signals resulting from V1 random patterns at the spatial scale of cortical columns

To simulate columnar-level MEG signals in human, we extracted the V1 cortical surface of a subject using Freesurfer automatic segmentation (Dale et al., 1999) (Fig. 6A). The triangulated surface had an average node distance of 880 μm (std = 279 μm), comparable to the diameter of orientation columns in human (Yacoub et al., 2008). Each node represented the center of a cortical column and was assigned random electrical activity sampled from a uniform distribution in the range of 0 and 1. The columnar-level simulated V1 activity was then mapped to 306 MEG sensors using a single sphere head model in Brainstorm (Tadel et al., 2011) (Fig. 6B).

To match the simulated MEG patterns in scale to an empirically realistic value of mean peak-to-peak strength of 978 femtoTesla (fT) as observed in experiment 1 (across-subject average of maximal trial-averaged evoked responses to grating stimuli), we scaled V1 pattern activity appropriately. This yielded V1 activation patterns in the range of 0 to 0.16 nano-Ampere-meter (nA-m) per simulated column. This is a physiologically plausible value, below the maximum current dipole of 18 nA-m per orientation column the V1 cortex can support, as derived from empirical and modeling studies. In particular, the basic unit of the neocortex is the minicolumn, with the primary visual cortex of macaque monkey having a density of approximately 1270 microcolumns per mm$^2$, and each microcolumn containing approximately 142 pyramidal cells (Jones, 2000). An orientation column in human, comprising several microcolumns, is about 800μm in diameter (Yacoub et al., 2008), and assuming the same cell density as the macaque monkey, it should contain approximately 90,000 pyramidal cells. Computational models (Murakami and Okada, 2006) and CA1 hippocampal pyramidal neuron measurements (Kyuhou and Okada, 1993) estimated the electrical activity of pyramidal neurons in the order of 0.2 pico-Ampere-meter per cell. This suggests a maximum current dipole of 18 nA-m per orientation column if all neurons activated synchronously and had dendrites parallel one another. Thus, the estimated maximum of 0.16nA-m per orientation columns is far below the expected electrophysiological maximum and can account for asynchronous neuronal firing, silent neurons, non-aligned dendrites, and other factors that would reduce the current dipole strength.

We estimated empirical noise from single trial baseline MEG data from Experiment 1 (average s.d. across subjects: 243 fT for magnetometers, 56.65 fT/cm for gradiometers). Adding noise at this level produced simulated MEG sensor patterns with signal-to-noise ratio (SNR) of -4 dB. To further explore a wide range of noise conditions, we added noise at different levels, ranging from -44 to 16 dB.

For every noise level, we evaluated whether simulated V1 activation patterns were discriminated by noisy MEG sensor patterns. We ran a multivariate analysis equivalent to the 'any phase' classification analysis of Experiment 1. For this, we simulated 500 pairs of random V1 activation patterns. For every pair, we created 960 noisy trials of MEG sensor level patterns for each V1 pattern. We sub-averaged simulated raw trials in groups of 40, and classified V1 patterns from averaged simulated MEG patterns. We repeated this process 100 times for random ascriptions of raw to averaged trials. Results (percent decoding accuracy) were averaged across the 100 iterations and the 500 random V1 activation patterns, yielding one decoding accuracy value for each noise level.

## 2.7 Statistical analysis

We used permutation tests for cluster-size inference, and bootstrap tests to determine confidence intervals for onset and peak latency of significant clusters (Nichols and Holmes, 2002; Pantazis et al., 2005; Maris and Oostenveld, 2007). Permutation tests exchanged the data labels (for example +45° vs. -45° grating orientation) randomly for each participant to determine significant time points of classification accuracy (10000 permutation samples, cluster-definition threshold p<0.05, cluster threshold p<0.05). Bootstrap tests sampled with replacement (1000 samples) the participant pool to estimate the distribution of onset and peak latency of significant clusters and derive 95% confidence intervals.

## 3 RESULTS

## 3.1 Experiment 1: Edge orientation in Cartesian gratings is decodable from MEG signals

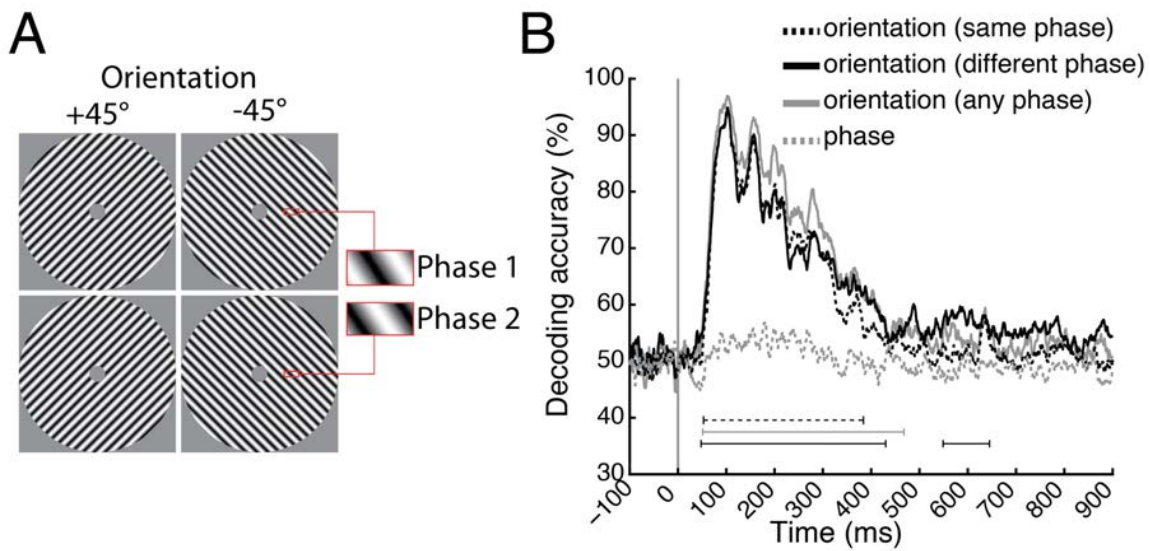

**FIGURE 1: Orientation decoding of oblique gratings. A)** The stimulus set comprised oblique gratings tilted right or left (orientation +/–45°) with two phase exemplars having a half cycle phase shift. **B)** Time course of orientation decoding in 3 cases: the classifier training and testing sets comprised grating stimuli of the same phase, different phase, or any phase. Grating orientation was robustly decoded in all analyses (also see Table 1A). There was no evidence for the representation of phase (classifier training and testing sets had the same orientation). Gray vertical line indicates stimulus onset. Lines below plots indicate

significant time points, color-coded as in decoding curves. (N=12; p<0.05 cluster definition threshold, p<0.05 cluster threshold).

The aim of experiment 1 was to establish whether edge orientation of Cartesian gratings is discriminated by visual representations measured with MEG. Previous studies have reported that cardinal orientations (i.e. 0 and 90°) were encoded differently from other orientations, either by a larger cell population (Pettigrew et al., 1968; Maffei and Campbell, 1970; Mansfield, 1974) or with sharper tuning curves (Rose and Blakemore, 1974; Kennedy and Orban, 1979). Thus, comparing brain responses to different cardinal orientations, or orientations of unequal angular disparity from cardinal orientations, might result in coarse-scale activation difference favoring cardinal orientations (cardinal bias). To avoid the coarse-scale cardinal bias, we used oblique gratings (+/–45°) whose orientation is equidistant from cardinal orientations (0/90°). We recorded MEG signals from 306 channels while 12 participants viewed a sequence of static Cartesian gratings (stimulus duration 100ms) in random succession (Fig. 1A). Gratings had two phase exemplars with a half cycle phase shift.

We extracted and preprocessed peri-stimulus MEG data from -100 to +900ms (1ms resolution) with respect to stimulus onset. We then used time-resolved multivariate pattern classification to decode the orientation of grating stimuli from MEG activation patterns. For each time point, MEG data were divided into training and testing sets, and the classifier (a linear support vector machine) learned to infer the orientation of grating stimuli from the training set. Decoding accuracy then quantified the performance of the classifier to predict the orientation of grating stimuli in the testing set. We determined statistical significance by non-parametric sign-permutation tests, and cluster-size inference for multiple comparison corrections (cluster-definition threshold p<0.05, cluster threshold p<0.05). Onset and peak latency of decoding time series are reported with 95% confidence intervals in brackets.

Restricting the training and testing sets to include grating stimuli of same phase, we found that MEG signals contained information about grating orientation starting at 53ms after stimulus onset (95% confidence interval 50-61ms), with a peak at 103ms (91-153ms). In contrast, an analogous decoding procedure for phase, with training and testing sets comprising grating stimuli of same orientation, did not reveal any phase information (Fig. 1B).

In the above analysis, grating stimuli that differed in orientation also consistently differed in local luminance because the phase was kept the same. Thus, orientation decoding may be due to local luminance differences rather than orientation. To determine whether grating orientation can be discriminated by visual representations independent of local luminance, we conducted a cross-classification analysis, forcing an assignment of opposite (half a cycle different) phases to the training and testing sets. In this analysis local luminance carries no information about orientation. We found that MEG signals resolved grating orientation independent of phase starting at 51ms (46-55ms), with a peak at 102ms (89-165ms), confirming that our results cannot be explained by local luminance differences.

Finally, we dropped phase information and assigned all grating stimuli to training and testing sets irrespective of phase. Such an approach affords higher signal-to-noise ratio by doubling the number of available images for the training and testing sets. We found that overall classification performance improved slightly, with an onset at 48ms (34-52ms) and a peak at 102ms (92-157ms).

In sum, we found that grating orientation was linearly decodable close to ceiling performance independent of phase for oblique (+/–45°) stimuli. We showed the time course with which orientation information is encoded by the visual brain (Garcia et al., 2013; Ramkumar et al., 2013), and excluded the coarse-scale effects of cardinal orientations as a necessary condition for orientation decoding.

## 3.2 Experiment 2: Edge orientation is decodable from MEG signals equally well for cardinal and oblique gratings

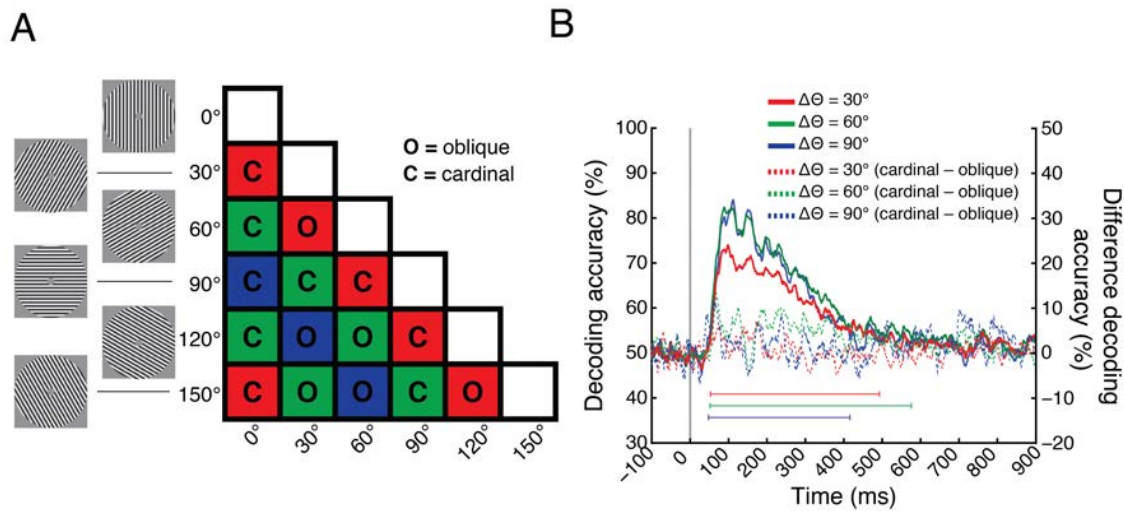

**FIGURE 2: Orientation decoding of cardinal and oblique gratings. A)** The stimulus set comprised Cartesian gratings with orientation 0° to 150° in steps of 30°, and had two phase exemplars with a half

cycle phase shift (not shown), as in Experiment 1. Pair-wise classifications were grouped by 1) orientation angle disparity Δθ (30°, 60° and 90°, color-coded in red, green and blue, respectively), and 2) the presence of at least one cardinal (C) or only oblique (O) orientations. **B)** Time-resolved orientation decoding for each Δθ (solid lines, left y-axis), and difference in orientation decoding between cardinal minus oblique cases for each Δθ (dotted line, right y-axis). Orientation was robustly decoded for all angle disparities (also see Table 1B). There was no evidence for stronger decoding of cardinal over oblique orientations. Gray vertical line and lines below plots same as in Figure 1(N=16, $p<0.05$ cluster definition threshold, $p<0.05$ cluster threshold).

Experiment 1 showed that coarse-scale activation differences through stimuli of cardinal orientations are not necessary for orientation decoding. However, it did not address the potential contribution of coarse-scale effects when cardinal orientations are being decoded. If cardinal orientations are encoded differently, this could result in MEG patterns favoring classification of cardinal orientations.

In experiment 2, Cartesian gratings with two cardinal angles (horizontal 0° and vertical 90°) and four oblique angles (30°, 60°, 120°, and 150°) were presented to 16 participants. We performed time-resolved orientation decoding across all possible pairs of gratings (Fig. 2A).

We first determined whether MEG data allows discrimination of grating orientations different by less than 90° for any phase. We found that orientation differences even as low as 30° were decodable (solid lines in Fig 2b; onset and peak latency in Table 1B). We then compared classification performance for the cases when decoding involved oblique gratings only, versus at least one grating with cardinal orientation (denoted 'o' and 'c' in Fig. 2A respectively). By subtracting decoding accuracy of oblique from cardinal cases, we found no evidence for differential encoding between oblique and cardinal orientations (Fig. 2B, dotted lines, right y-axis). Note that equivalent results were obtained in a cross-classification analysis by assigning opposite phases to the training and testing sets (Table 1B).

In sum, the results of experiment 2 complemented the ones of experiment 1 in two ways. First, they showed that MEG signals contain orientation information at least as low as 30°. Second, by failing to provide evidence for differential encoding of cardinal and oblique orientations, they suggest that even for cardinal orientations the role of the cardinal bias in decoding orientation is small, if any.

## 3.3 Experiment 3: Edge orientation in radially balanced logarithmic spirals is decodable from MEG signals

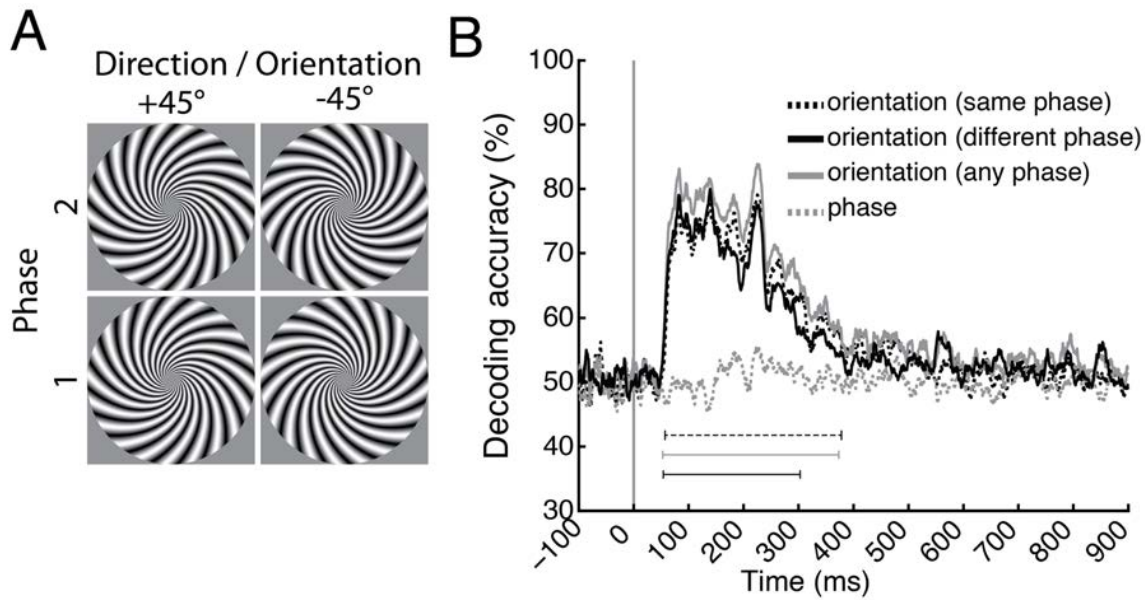

**FIGURE 3: Orientation decoding of radially balanced stimuli. A)** The stimulus set comprised four exponential spirals in two opposing directions and two phase exemplars having a half cycle phase shift in phase. At any point, the stimuli of opposing directions were orthogonal to each other (90° disparity). Importantly, at any point the disparity of spiral orientation to a radial line was 45 ° (+/− for the different directions), thus balancing the radial component. **B)** Time course of orientation decoding in three cases: the classifier training and testing sets comprised grating stimuli of the same phase, different phase, or any phase. Spiral direction (orientation) was robustly decoded in all analyses (for details see Table 1C). There was no evidence for the representation of phase (classifier training and testing sets had the same orientation)**.** Gray vertical line and lines below plots same as in Figure 1. (N=13, p<0.05 cluster definition threshold, p<0.05 cluster threshold).

Experiment 3 was designed to address a second potential orientation-specific coarse-scale bias: the radial bias. The radial bias is the differential representation of orientations collinear with a line through the point of fixation (radial line) (Sasaki et al., 2006; Mannion et al., 2009; Freeman et al., 2011; Alink et al., 2013). For example, in the above experiments oblique +45° gratings might activate stronger cortical regions representing the upper right and lower left rather than the upper left and lower right quadrants of the visual field. Oblique -45° gratings would have an opposite bias.

To control for radial bias, we presented participants with radially balanced logarithmic spirals (Mannion et al., 2009) turning in two opposing directions (Fig 3A). The stimuli were designed such that at any point spirals of opposing direction are oriented orthogonal to each other, just as Cartesian gratings of 90° disparity. Note that the disparity of exponential spirals with respect to a radial line is exactly 45° (+ or – depending on the turning direction), thus balancing the radial component of the stimulus.

We conducted multivariate pattern classification as in experiment 1, by restricting the training and testing sets to have grating stimuli of same phase, different phase, or any phase. Results were similar as above: robust decoding of orientation in all analyses, and no evidence for decoding of phase (Fig. 3B; Table 1C).

In sum, experiment 3 showed that the orientation of radially balanced stimuli was robustly discriminated by visual representations, excluding a coarse-scale retinotopic bias as a likely source of orientation signals observed with MEG.

## 3.4 Experiment 4: Edge orientation is decodable from MEG signals independent of grating shape and thus stimulus edge effects

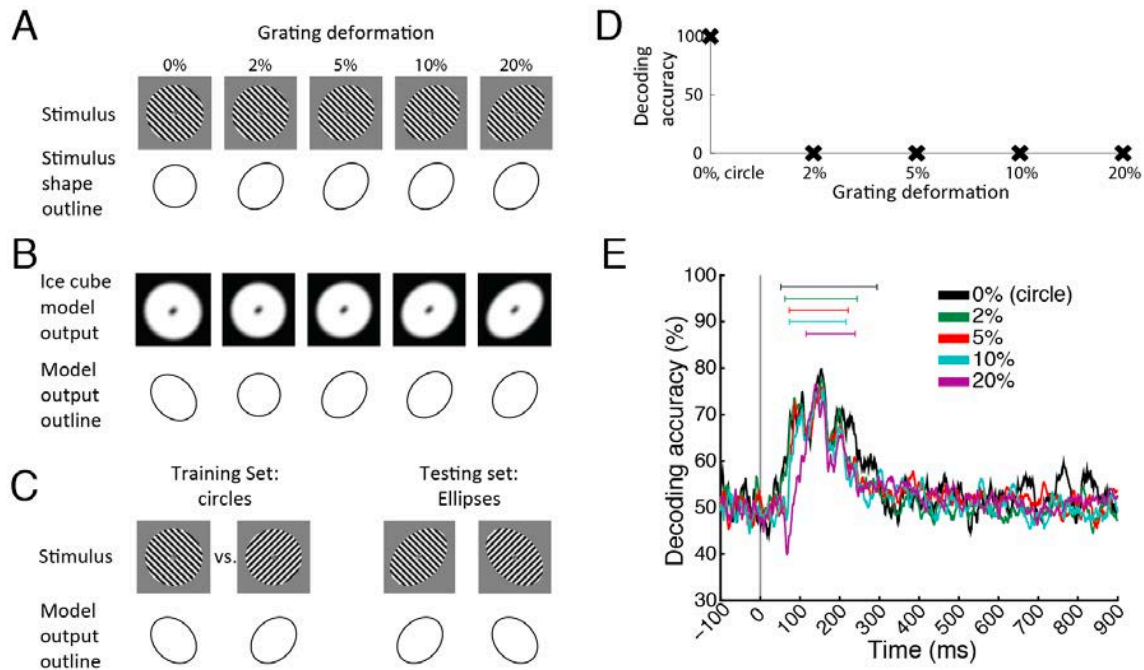

**FIGURE 4: Orientation decoding of gratings with different shape outlines. A)** The stimulus set comprised gratings +/–45° (here shown –45°). The type of stimulus shape outline (circle or ellipse) is indicated with a sketch below each stimulus. **B)** Ice cube model output for the above experimental stimuli. The ice cube model produced outputs with subtle edge artifacts: elongation in the direction of the grating,

and compression in the orthogonal direction. As a result, stimuli with a shape outline of a circle produced a model output with a shape outline of an ellipse. **C)** Experimental approach to evaluate the influence of edge artifacts on orientation decoding. A classifier is trained to distinguish orientation based on grating stimuli of a circular outline shape (but ellipsoidal model outline shape), and tested on grating stimuli with an ellipsoidal outline shape. If the model output is relevant in decoding (deformation by edge effects), the classifier will have poor performance because the shape of model output conflicts with orientations**. D)** Decoding results on ice-cube model output. We found that for shape deformations of 2% or more, the classifier predicted orientation incorrectly. Thus, if the ice cube model is correct and if edge effects determine empirical orientation decoding of gratings, classification of orientation from MEG data across grating shapes should result in near 0 decoding accuracies. In contrast, high decoding accuracies would indicate that local orientation is the relevant factor, and edge artifacts are unlikely to explain orientation decoding from MEG data. **E)** Classification of MEG responses indicated neural representations robust even to large changes in grating outline shape. Gray vertical line and lines above plots same as in Figure 1. (N=12, p<0.05 cluster definition threshold, p<0.05 cluster threshold).

Grating stimuli are spatially limited with annulus boundaries that induce edge effects dependent on the orientation of the stimulus. In particular, a perfect ice-cube model of primary visual cortex (Hubel and Wiesel, 1959, 1968) with no bias in the number of neurons representing different orientations can still account for orientation decoding (Carlson, 2014). This is because the representation of an exact circular grating is in fact an ellipse elongated in the direction of the grating, and compressed in the orthogonal direction. Thus, signals differentiating the orientation of stimuli might actually originate from the differential location of those edge effects (see simulations in Matlab language in Text S1). To evaluate the influence of such edge effects, we created a stimulus set of Cartesian gratings shaped as ellipses elongated in the direction opposite to the one predicted by the model (Fig 4A). If the model prediction is valid, training a classifier to discriminate orientation in circular and testing on ellipsoidal gratings should significantly compromise the classifier performance. If however the source of orientation signals in MEG is independent of edge effects, the classifier should correctly predict the orientation even for heavily distorted grating shapes.

Based on our modeling simulations (Fig. 4A-D), grating stimuli distorted above ~2% should result in opposite orientation decoding if the model accurately predicts V1 decoding. Therefore, our stimulus set consisted of phase-randomized gratings of +/–45° orientation in shapes ranging from perfect circles to ellipses of 2, 5, 10 and 20% distortion (Fig 4A). We then trained a classifier to distinguish orientations for circular gratings, and tested on ellipsoidal gratings (Fig. 4E, Table 1D). Training and testing on circular gratings provided a baseline control. We observed that even very strong deformations of the annulus by as much as 20% of the radius did not compromise robust orientation classification, countering the prediction of the perfect ice-cube model.

These results indicate that coarse-scale edge effects as predicted by the ice-cube model are not necessary for robust orientation decoding, and that orientation signals can be robustly read out from MEG data independent of the overall shape of the stimulus.

## 3.5 Experiment 5: Edge orientation in stimuli with similar global form is decodable from MEG signals

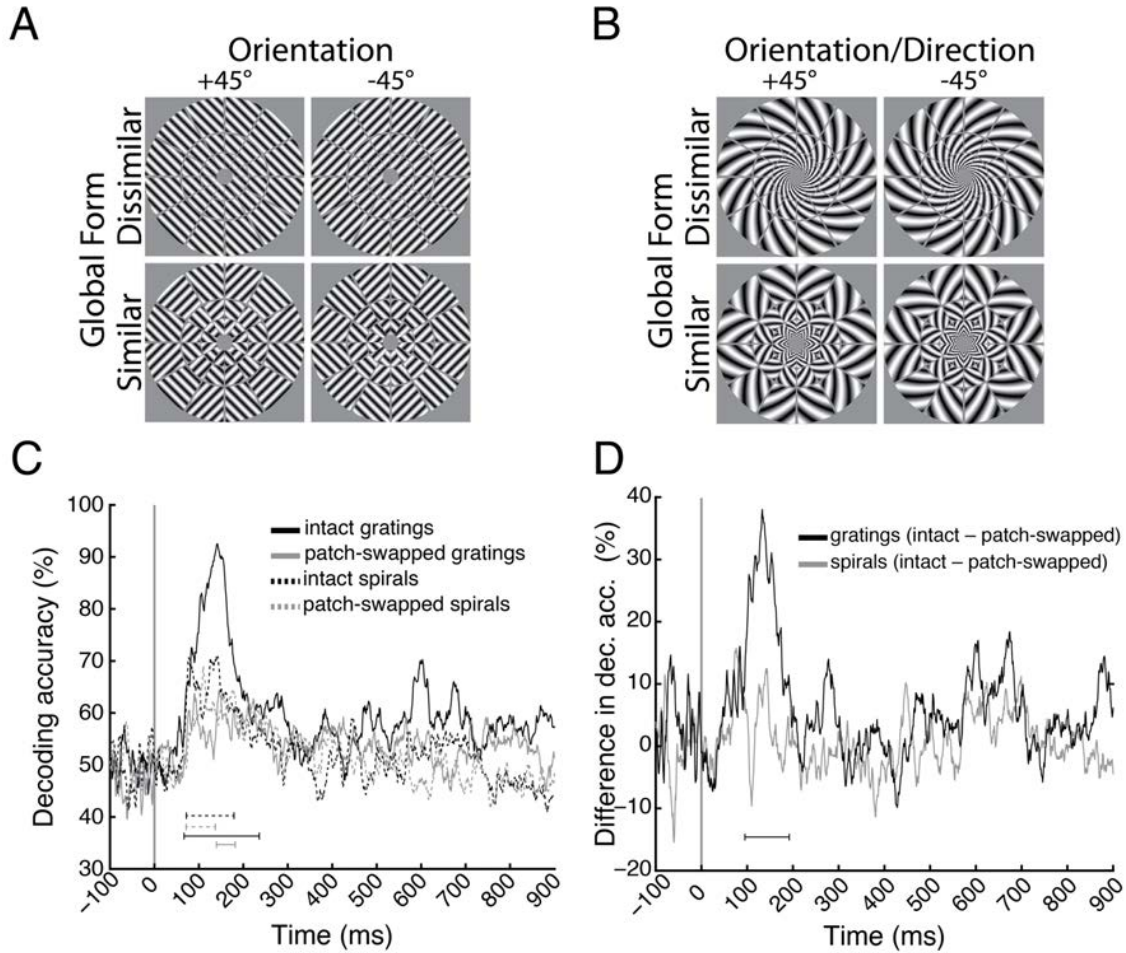

**FIGURE 5: Influence of global form on orientation encoding. A,B)** The stimulus set comprised oblique gratings (same as experiment 1) and radially balanced exponential spirals (same as experiment 3) in both intact and patch-swapped forms. Whereas intact stimuli differed both in local orientation and global form, patch-swapped stimuli differed only in local orientation. **C)** Time course of orientation decoding for intact and patch-swapped gratings and spirals. Orientation was discriminated by visual representations in all conditions (for details see Table 1E). **D)** Comparison of classification between intact and patch-swapped stimuli. Gray vertical lines and lines below plots same as in Figure 1. (N = 12, p<0.05 cluster definition threshold, p<0.05 cluster threshold).

Gratings and spirals are coherent stimuli that differ not only in local orientation, but also in global form. For example, an intact grating at +/–45° may look like a coherent object rotated to the left or right, eliciting global form related signals at a coarse-scale level (Alink et al., 2013). To dissociate local orientation from global form, it is necessary to compare brain responses to stimuli that differ in local orientation but not global form. Such stimuli can be created by patch-swapping intact stimuli (gratings or spirals) in non-adjacent regions defined by a polar checkerboard array, as in (Alink et al., 2013).

Experiment 5 presented intact and patch-swapped gratings and spirals to 12 participants (Fig. 5A). Stimuli were the same as in Experiments 1 and 3, and their patch-swapped variants. To control for the additional edges introduced by the swapping, lines of background color covered the patch edges for all stimuli. Classifiers were trained to discriminate orientation of all stimuli irrespective of phase (any phase, Fig. 5C).

Corroborating experiments 1-4, we found that MEG signals contained information about orientation of intact gratings and spirals (Fig. 5C, black curves, solid for spirals, dotted for gratings; for details on onsets of significance and peaks see Table 1E). Crucially, MEG signals contained information about patch-swapped variants of spirals and gratings as well (Fig. 5C, gray curves, solid for spirals, dotted for gratings). This indicates that global form differences are not necessary for decoding of orientation from MEG signals.

Comparing decoding accuracy of patch-swapped with intact stimuli, we found a significant decrease for gratings with an onset at 95ms (80-117ms) and a peak at 133ms (116-141ms), but no difference for spirals (Fig. 5D, black curve for gratings, gray for spirals; for details see Table1E).

In sum, these results indicate that global form influences orientation encoding in cortex, but is not necessary for orientation decoding. They further reveal the time at which global form first influences orientation encoding for grating stimuli.

## 3.6 Experiment 6: Simulated neuronal activation patterns at a spatial scale comparable to orientation columns are decodable from modeled MEG signals under realistic noise conditions

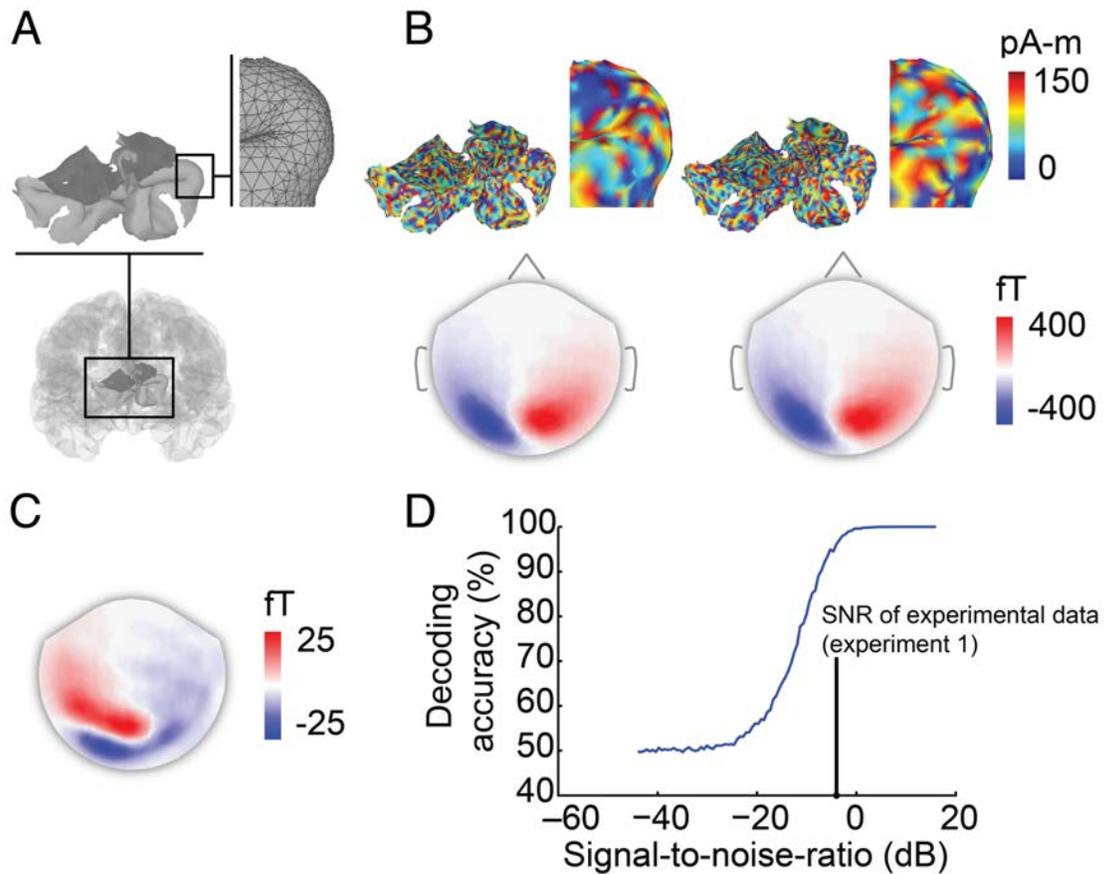

**FIGURE 6: Decoding of simulated random patterns differing at columnar-level spatial scale in human V1 from MEG sensor data**. A) Refined triangulated mesh of V1 cortex with an average node distance of 880 μm (s.d. = 279 μm) comparable to the diameter of orientation columns in human. **B**) Two example random activation patterns of simulated columnar-level neuronal activity in V1 (top), and topography of the corresponding MEG measurements (bottom). **C)** Difference between the two MEG topographies shown in B. **D)** Decoding random activity patterns of V1 activity at different SNR levels. Decoding accuracy was 96.1% when SNR was equal to experimental data, and decayed gradually for lower SNR levels.

We conducted a simulation experiment to test whether the physics of MEG can support the conjecture that orientation information may originate from V1 cortical patterns at the spatial scale of orientation columns. MEG has a coarse spatial resolution, which is highly non-uniform and in the order of several millimeters on the visual cortex (Hämäläinen et al., 1993; Darvas et al., 2004). The elemental model sources of MEG signals are current dipoles oriented normally to the cortex and neighboring current dipoles are too close to be resolved if they are parallel to one another. However, due to the highly convoluted nature of the cortical manifold, neighboring current dipoles are rotated in space thus producing distinct magnetic fields. Two activation patterns on cortex that differ on the scale of orientation columns would therefore give rise to different magnetic field topographies outside the head. Orientation information represented on the scale of orientation columns might as a result be resolvable with multivariate MEG methods given the complex folding pattern of cortex.

To test this hypothesis, we evaluated whether simulated random activation patterns in V1, differing at the spatial scale of columns, could be discriminated by the corresponding

MEG sensor patterns when signal strength and noise were equated to empirically measured data. The simulation was based on a triangulated surface of the V1 cortical sheet with an average node distance of 880 μm (s.d. = 279 μm), comparable to the diameter of orientation columns in human (Yacoub et al., 2008) (Fig. 6A). We generated neuronal activation patterns on the V1 surface by assigning random activation values to each node (Fig. 6B), and then computed the corresponding MEG sensor level signals by forward modeling. These simulated MEG measurements had overall similar topographies, but with weak differences (Fig. 6BC). The data was scaled to match the range of measured data, and sensor-level Gaussian noise was then added at various signal-to-noise ratios (SNRs).

We used multivariate pattern classification to discriminate different V1 activation patterns from the noisy instantiations of MEG sensor signals. We found that simulated V1 patterns were discriminated with 96.1% accuracy from MEG sensor patterns when adding sensor-level noise at the same SNR level as the experimental data (Figure 6D). For lower SNR levels, decoding accuracy declined gradually.

These modeling results demonstrate the theoretical feasibility of discriminating activation patterns in V1, differing at the spatial scale of cortical columns, from MEG signals.

## 4   DISCUSSION

We investigated the bound on the spatial scale of neuronal activation patterns from which information can be decoded from MEG signals. Our analysis focused on decoding of contrast edge orientation, a visual feature known to be encoded in fine-scale neuronal activation patterns in V1. In five empirical experiments we found that MEG signals contained orientation information even when controlling for factors known to induce coarse-scale activation in V1, such as overrepresentation of particular orientations, boundary interaction effects between background and stimulus, and global form-related signals. Theoretical modeling demonstrated the feasibility of discriminating signals at the spatial scale of orientation columns from MEG data under realistic noise conditions. Taken together, our results show that neural signals at the level of cortical orientation columns are accessible by electrophysiological measurements in humans, and likely subserve orientation decoding. This result has wide implications for the interpretation of studies using MEG and multivariate pattern classification, suggesting that any type of information encoded in the human brain at the level of cortical columns should be accessible by MEG.

## 4.1 Non-invasive electrophysiological signals in humans carry information about orientation

Most recent studies investigating orientation encoding in the human brain have used fMRI (Haynes and Rees, 2005; Kamitani and Tong, 2005; Mannion et al., 2009, 2010; Kriegeskorte et al., 2010; Freeman et al., 2011, 2013; Alink et al., 2013; Carlson, 2014; Carlson et al., 2014; Pratte et al., 2014), and only a few studies used non-invasive electrophysiological methods (Campbell and Maffei, 1970; Duncan et al., 2010; Koelewijn et al., 2011; Garcia et al., 2013). However, using an electrophysiological method such as MEG offers new insights not possible with fMRI. MEG is not confounded by interpretative complications posed by the complex relationship between neural activity, BOLD contrast, and voxel sampling (Haynes and Rees, 2005; Kamitani and Tong, 2005; Mannion et al., 2009, 2010; Kriegeskorte et al., 2010; Freeman et al., 2011, 2013; Alink et al., 2013; Carlson, 2014; Carlson et al., 2014; Pratte et al., 2014). Importantly, it allows resolving the precise timing of orientation encoding in the order of milliseconds, impossible with fMRI due to the sluggishness of the BOLD response. This allows timing-based dissociation of first-pass responses in early visual regions from possibly more distributed responses that may involve feedback information. Our study decoded orientation directly from neural activity in millisecond resolution.

Previous electrophysiological studies investigating orientation decoding used adaptation (Campbell and Maffei, 1970) or decoding in frequency space (Duncan et al., 2010; Koelewijn et al., 2011; Garcia et al., 2013), providing a temporal resolution of 25ms at best. Recently, a study employed multivariate pattern classification to resolve the time course of orientation encoding of Cartesian gratings in millisecond resolution (Ramkumar et al., 2013). Our experiments build upon those studies by clarifying the nature of the sources of orientation-selective signals in noninvasive electrophysiological methods. By controlling for possible stimulus confounds that could induce coarse-scale activation patterns in V1, our results indicate that orientation information in MEG signals originates from the spatial scale of cortical columns.

Note that our analysis framework does not allow us to unequivocally localize signal sources to V1 only, and not to other cortical areas such as V2 and V3 (Haynes and Rees, 2005; Kamitani and Tong, 2005) that have been shown to encode orientation or global form (Ostwald et al., 2008; Seymour et al., 2010). However, we observed orientation-selective MEG signals starting already ~50ms after stimulus onset. This short latency is consistent with the latency of V1 spike time in monkeys (Schmolesky et al., 1998; Bullier, 2001; Mormann et al., 2008) and the C1 component in the visual event-related brain potential (Jeffreys and Axford, 1972; Clark et al., 1994; Russo et al., 2003), and inconsistent with a large contribution of other cortical areas with longer spiking latencies, as well as feedback processing. This suggests that the observed MEG signals, at least in their early phase, likely originate in V1. Further studies that would localize MEG components to V1 based on subject-specific cortical surfaces reconstructed from MRI might yield corroborative evidence.

Complementing the experimental results, our modeling experiment demonstrated the theoretical feasibility of discriminating activation patterns in V1 differing at the spatial scale of cortical columns from MEG signals. Despite the low spatial resolution of MEG, the highly folded V1 cortex around the calcarine fissure gave rise to distinct and decodable magnetic fields outside the head. Conventional MEG modeling approaches consider either a small number of focal cortical sources, or cortically distributed source models with extended but similar activity in cortical patches (Baillet et al., 2001). Our results show that distributed sources of high spatial structure are also resolvable with MEG in folded areas of the cortex.

In total, our results indicate that MEG potentially can distinguish edge orientations based on signals originating at the level of orientation columns. This sets the stage for a direct investigation of edge orientation in the human brain, a fundamental visual property.

Further, our results have implications for the interpretation of MEG studies using multivariate pattern classification in other visual contents, such as objects (Carlson et al., 2013; Cichy et al., 2014; Isik et al., 2014). Although for complex objects coarse-scale activation pattern differences are expected (Op de Beeck et al., 2008), a contribution of signals at the level of cortical columns (Fujita et al., 1992; Wang et al., 1996) should also be considered.

### 4.2 How strong is the cardinal and radial bias?

When stimuli are not balanced in their cardinal or radial components, how strong can this influence orientation decoding from MEG data? Concerning the cardinal bias, experiment 2 unexpectedly did not reveal any differences in decoding for cardinal versus non-cardinal orientations. This suggests that the cardinal bias has weak influence in orientation decoding. Note however that existing studies on the neural basis of the cardinal effect have been mixed. Using fMRI and MEG in humans, some studies reported stronger (Zemon et al., 1983; Moskowitz and Sokol, 1985; Furmanski and Engel, 2000; Yang et al., 2012), others reduced (Serences et al., 2009; Mannion et al., 2010; Swisher et al., 2010) responses to cardinal orientations, or both depending on timing (Koelewijn et al., 2011). While many studies have found or postulated larger population sizes for neurons tuned to cardinal orientations (21–23, 45–49), others have indicated large variability across subjects (Chapman and Bonhoeffer, 1998). Finally, the behavioral effect itself is complex and dependent on factors, such as context of the oriented edges and eccentricity (Essock et al., 2003, 2009; Westheimer, 2003, 2005; Hansen and Essock, 2004), suggesting that the neural effect may also be strongly dependent on particular stimulus parameters.

Concerning the radial bias, findings from experiment 3 are consistent with a possible contribution of coarse-scale radial bias signals to orientation decoding: classification accuracy for gratings (radially unbalanced) was higher than for spirals (radially balanced) in all experiments (Sasaki et al., 2006; Carlson et al., 2013). A direct comparison of gratings and spirals however is not possible, as radial gratings change stripe width with eccentricity, while Cartesian gratings do not. Differences in decoding accuracy might instead reflect cortical differences in spatial frequency sensitivity varying with eccentricity (Sasaki et al., 2006). Finally, a recent fMRI study observed coarse-scale biases for spirals of different turning direction, questioning the validity of logarithmic spirals as a proper control for coarse-scale activation biases (Freeman et al., 2013). However, the reported results may be influenced by edge-related artefacts (Carlson, 2014), and may be crucially dependent on particular analysis choices (Pratte et al., 2014). In general, future MEG studies that address any newly observed coarse-scale biases by proper stimulus material will be necessary to test the relative contributions of fine- and coarse-scale sources of orientation decoding.

### 4.3 Orientation signals in MEG are not explained by edge-related effects

Previous research has suggested that interaction effects at the boundary between background and oriented grating stimuli might carry information about orientation (Carlson, 2014). Controlling for these effects in experiment 4 did not abolish robust orientation decoding from MEG signals. Thus, although edge related effects might have biased previous studies, they are not necessary for orientation decoding from MEG signals, and unlikely to produce large effects for commonly used stimulus configurations.

## 4.4 Global form influences orientation encoding, but is not necessary for orientation decoding

What is the source of the global form effects on orientation encoding? An fMRI study that found global form influences in orientation encoding in early visual cortex proposed three possibilities (Alink et al., 2013): a) feedback from inferior temporal cortex, b) attentional spread along Gestalt criteria, and c) contextual modulation effects. Our MEG results can tentatively differentiate between those alternatives by timing: global form effects can originate only from a neuronal process with an earlier onset, and should have comparable peak latency. Intracranial recordings in inferior temporal cortex have shown that global form modulates neural activity with short onset latency (~80ms), but with late peak responses at ~200ms (Brincat and Connor, 2006), which is at odds with the earlier peak latency observed here. Attentional spread along Gestalt criteria has been observed in V1 with an onset latency of approximately 330ms (Wannig et al., 2011), strongly inconsistent with the onset of global form effects. However, contextual modulation effects in V1, which occur with an onset latency of 100ms and peak latencies of 150ms (Lamme, 1995; Zipser et al., 1996), are consistent in timing with our findings.

## 4.5 Conclusions

We found that MEG signals allowed decoding of contrast edge orientation as early as ~50ms. Importantly, corroborating evidence from 5 experiments indicated that this information originates from spatially fine patterns in orientation columns, since decoding was possible even when controlling for multiple known confounds known to induce coarse-scale activation in V1, such as overrepresentation of particular orientations, boundary interaction effects between background and stimulus, and global form-related signals. Our V1 modeling study further demonstrated the feasibility of decoding information encoded at the spatial scale of cortical columns. Generalizing from this evidence, any information encoded in the human brain at the level of cortical columns, and not only contrast edge orientation, should in principle also be accessible by multivariate analysis of electrophysiological signals.

## 5 ACKNOWLEDGEMENTS

This work was funded by the National Science Foundation (BCS-1134780 to D.P.), the Humboldt Foundation (Feodor Lynen Fellowship to R.M.C), and was conducted at the Athinoula A. Martinos Imaging Center at the McGovern Institute for Brain Research, Massachusetts Institute of Technology. We thank Thomas Christophel, Charles Jennings, Seyed-Mahdi Khaligh-Razavi and Santani Teng for comments on the manuscript.

# 7 TABLES

| Decoding Analysis | Onset latency (ms) | Peak latency (ms) |
|---|---|---|
| **A) Experiment 1** | | |
| Grating orientation (same phase) | 53 (50-61) | 103 (91-153) |
| Grating phase (same orientation) | - | - |
| Grating orientation (different phase) | 51 (46-55) | 102 (89-165) |
| Grating orientation (any phase) | 48 (34-52) | 102 (92-157) |
| **B) Experiment 2** | | |
| Grating orientation at 30° disparity (any phase) | 52 (44-55) | 99 (82-158) |
| Grating orientation at 60° disparity (any phase) | 52 (45-55) | 89 (84-149) |
| Grating orientation at 90° disparity (any phase) | 48 (42-54) | 112 (88-156) |
| Grating orientation at 30° disparity (different phase) | 60 (54-65) | 90 (70-156) |
| Grating orientation at 60° disparity (different phase) | 56 (49-62) | 93 (82-152) |
| Grating orientation at 90° disparity (different phase) | 57 (52-59) | 104 (82-152) |
| **C) Experiment 3** | | |
| Spiral orientation (same phase) | 57 (52-59) | 225 (80-230) |
| Spiral phase (same orientation) | | |
| Spiral orientation (different phase) | 54 (48-56) | 139 (82-224). |
| Spiral orientation (any phase) | 53 (49-56) | 225 (82-230) |
| **D) Experiment 4** | | |
| Grating orientation (circular annulus) | 52 (39-134) | 154 (98–158) |
| Grating orientation (across circular and ellipsoidal annulus (2%)) | 73 (69-74) | 154 (87–161) |
| Grating orientation (across circular and ellipsoidal annulus (5%)) | 50 (44-73) | 155 (86-158) |
| Grating orientation (across circular and ellipsoidal annulus (10%)) | 73 (71-124) | 158 (96-160) |
| Grating orientation (across circular and ellipsoidal annulus (20%)) | 112 (90–135) | 140 (115-204) |
| **E) Experiment 5** | | |
| Orientation of intact gratings | 69 (52-88) | 142 (132-155) |
| Orientation of patch-swapped gratings | 142 (72-641) | 145 (78-746) |
| Orientation of intact spirals | 72 (60-126) | 139 (76-143) |
| Orientation of patch-swapped spirals | 83 (75-234) | 110 (109-235) |
| Difference between intact and patch-swapped gratings | 95 (80-117) | 133 (116-141) |
| Difference between intact and patch-swapped spirals | – | 76 (87 – 699) |

**Table 1: Onset and peak latencies of first significant cluster in experiments 1-5.**
Values are means across participants and 95% confidence intervals in brackets as determined by bootstrapping the participant pool (1000 times).